\documentclass[runningheads]{llncs}
\usepackage{hyperref}
\usepackage{subcaption}
\usepackage{tabularx}
\usepackage{listings}
\usepackage{verbatim}
\usepackage[table]{xcolor}
\usepackage{setspace}
\usepackage{enumitem}
\usepackage{float}
\usepackage[]{caption}
\usepackage{orcidlink}
\usepackage{pdfpages}
\usepackage{enumitem}
\usepackage{graphicx}
\usepackage{booktabs}
\usepackage{array}
\begin{document}
\title{PM4Py.LLM: a Comprehensive Module for Implementing PM on LLMs}
\titlerunning{PM4Py.LLM: a Comprehensive Module for Implementing PM on LLMs}

\author{
Alessandro Berti\inst{1,2}\orcidlink{0000-0002-3279-4795}
}
\authorrunning{A. Berti}
\institute{Process and Data Science Chair, RWTH Aachen University, Aachen, Germany \and
Fraunhofer FIT, Sankt Augustin, Germany  \\
\email{alessandro.berti@fit.fraunhofer.de}}

\maketitle

\begin{abstract}
\emph{pm4py} is a process mining library for Python implementing several process mining (PM) artifacts and algorithms.
It also offers methods to integrate PM with large language models (LLMs). This paper examines how the
current paradigms of PM on LLM are implemented in \emph{pm4py}, identifying challenges such as privacy, hallucinations,
and the context window limit.
\end{abstract}

\renewcommand{\sectionautorefname}{Section}
\renewcommand{\subsectionautorefname}{Section}
\renewcommand{\subsubsectionautorefname}{Section}

\section{Introduction}
\label{sec:introduction}

Process mining (PM) is a branch of data science aiming to extract process-related insights from the event data recorded by information systems supporting the execution of such processes.
Both open-source (ProM) and commercial (Fluxicon Disco, Celonis, SAP Signavio) tools are available for PM. Among open-source offerings, the \emph{pm4py} process mining library for Python \cite{DBLP:journals/simpa/BertiZS23}
is a popular option implementing various process mining techniques, including process discovery, conformance checking, and model enhancement.
\emph{pm4py}, being a library, implements several PM artifacts, including event log structures, process models (process trees, Petri nets, BPMN models), and machine learning features.
Large Language Models (LLMs) and Large Vision Language Models (LVLMs) have been recently implemented for PM tasks in different implementation paradigms \cite{DBLP:conf/bpm/Berti0A24}.
The goal of this paper is to present how \emph{pm4py}, allowing for the implementation of different paradigms, integrates with large language models.
In \cite{DBLP:conf/bpm/Berti0A23}, some textual abstractions of event logs and process models provided by \emph{pm4py} are described.
However, many more techniques (visual abstractions, APIs for code generation) are currently provided in the library, that are introduced in this paper and are important to extend the scope of PM on LLMs.
The rest of the paper is organized as follows. Section \ref{sec:implementationParadigms} describes the current paradigms for PM on LLMs; Section \ref{sec:implementation} describes the implementation of the paradigms in \emph{pm4py};
Section \ref{sec:conclusion} concludes the paper with an outlook over possible development directions.

\section{Paradigms for PM on LLMs}
\label{sec:implementationParadigms}

The following paradigms have been adopted for PM on LLMs \cite{DBLP:conf/bpm/Berti0A24}:
\begin{itemize}
\item \emph{Direct Provision of Insights}: A prompt containing abstractions of PM artifacts, domain knowledge, and questions is provided to the LLM/LVLM. Different techniques are available for the abstraction of PM artifacts:
\begin{itemize}
\item \emph{Textual Abstractions of Process Mining Artifacts}: The PM artifact is provided as text \cite{DBLP:conf/bpm/Berti0A23}.
\item \emph{Visual Abstractions of Process Mining Artifacts}: The PM artifact is provided as a visual representation, which can be interpreted by the LVLM.
\end{itemize}
Typical questions that are possible using this paradigm are: ``What are the anomalies in the event data?'', ``Can you explain the process?''.
The main limitation is the context window limit of the LLM/LVLM. Moreover, the proposed implementation paradigm might suffer from privacy (data needs to be provided to the LLM) and \emph{hallucinations} by the LLM/LVLM \cite{DBLP:conf/esws/MartinoIT23}.
In \emph{pm4py}, the textual/visual abstractions can be provided along with a question to an LLM. In particular, $\mathtt{pm4py.llm.openai\_query}$ is implemented, which allows querying OpenAI's LLMs/LVLMs (such as \emph{gpt-4-turbo-preview} and \emph{gpt-4-vision-preview}).
\item \emph{Translation of Natural Language Statements to SQL Queries}: Alternatively, the LLM can generate SQL queries executable on the process mining dataset \cite{DBLP:journals/corr/abs-2307-09909}. This mitigates the privacy and hallucination issues.
However, the user needs to provide a clear question and all the required domain knowledge on the dataset and on the process \cite{DBLP:journals/corr/abs-2307-09909}.
Typical questions that are possible using this paradigm are: ``How many cases are contained in this event log?'', ``Can you measure the average throughput time of the cases containing the activity X?''.
\item \emph{Code Generation}: Using the LLM to generate scripts executable against the process mining dataset is a paradigm applied successfully in process modeling \cite{DBLP:conf/bpm/Humam0A24}. It requires the detailed specification of the APIs to be used.
Typical statements that are possible using this paradigm are: ``Discover a process model from the data'', ``Extract an event log from my Outlook calendar''.
\item \emph{Automatic Formulation of Hypotheses}: This paradigm provides textual abstractions of the process mining artifacts in order for the LLM to generate hypotheses over the artifact \cite{DBLP:conf/bpm/Berti0A23}.
A SQL query is also provided to verify the validity of such hypotheses. Refinement cycles can be realized when the original hypotheses are not valid, leading to refined hypotheses.
\end{itemize}
The provision of the original process mining artifact is usually not a viable option due to the context window limit, as real-life event logs and process models have millions of events and hundreds of activities.

\section{Implementations}
\label{sec:implementation}

\begin{table*}[!t]
\centering
\resizebox{\textwidth}{!}{
\begin{tabular}{ll}
\toprule
\textbf{Command} & \textbf{Description} \\
\midrule
\texttt{pm4py.llm.abstract\_dfg} & Provides the DFG abstraction of a traditional event log. \\
\texttt{pm4py.llm.abstract\_variants} & Provides the variants abstraction of a traditional event log. \\
\texttt{pm4py.llm.abstract\_log\_attributes} & Provides the abstraction of the attributes/columns of the event log. \\
\texttt{pm4py.llm.abstract\_log\_features} & Provides the abstraction of the machine learning features obtained from an event log. \\
\texttt{pm4py.llm.abstract\_case} & Provides the abstraction of a case (collection of events). \\
\texttt{pm4py.llm.abstract\_ocel} & Provides the abstraction of an object-centric event log (list of events and objects). \\
\texttt{pm4py.llm.abstract\_ocel\_ocdfg} & Provides the abstraction of an object-centric event log (OC-DFG). \\
\texttt{pm4py.llm.abstract\_ocel\_features} & Provides the abstraction of an object-centric event log (features for ML). \\
\texttt{pm4py.llm.abstract\_event\_stream} & Provides an abstraction of the (last) events of the stream related to a traditional event log. \\
\texttt{pm4py.llm.abstract\_temporal\_profile} & Provides the abstraction of a temporal profile model. \\
\texttt{pm4py.llm.abstract\_petri\_net} & Provides the abstraction of a Petri net. \\
\texttt{pm4py.llm.abstract\_declare} & Provides the abstraction of a DECLARE model. \\
\texttt{pm4py.llm.abstract\_log\_skeleton} & Provides the abstraction of a log skeleton model. \\
\bottomrule
\end{tabular}
}
\caption{Textual abstractions of PM artifacts provided by \emph{pm4py}.}
\label{table:pm4py_llm_textual_abstractions}
\vspace{-4mm}
\end{table*}

In this section, the implementation in \emph{pm4py} of the different paradigms described in Section \ref{sec:implementationParadigms} is discussed.

\vspace{-2mm}
\subsection{Textual Abstractions of Process Mining Artifacts}
\label{subsec:textualAbstractionProcMinArt}

Textual abstractions of PM artifacts can be provided (up to the context window limit) to any LLM. The possible abstractions are described
in Table \ref{table:pm4py_llm_textual_abstractions}.
For traditional event logs, the top entries of the directly-follows graph (couples of activities following each other in at least a case of the log)
or the top process variants can be computed. Moreover, it is possible to extract a summary of the numerical machine learning features extracted from the event log (each feature is provided with different quantiles).
For object-centric event logs, the top entries of the object-centric directly-follows graph can be extracted. As for traditional event logs, it is possible to extract a summary of the machine learning features extracted from the object-centric event log.
It is also possible to abstract the entire list of events and objects of the object-centric event log. However, the resulting textual abstraction would probably exceed the context window limit of the LLM if not limited to a subset of events/objects.
The methods support the specification of the context window limit. The abstractions usually provide, at first, the most important entry and continue to include entries until the specified length limit is exceeded.

Petri nets can be textually abstracted as their set of places/transitions/arcs \cite{DBLP:conf/bpm/Berti0A23}. Declarative process models, such as DECLARE \cite{DBLP:conf/bpm/Maggi13} and the log skeleton \cite{DBLP:journals/sttt/Verbeek22}, can also be textually abstracted. The exact specification of the artifact can vary. However,
since LLMs might not have the knowledge of a specific declarative notation, the description of the rules of the notation should also be included in the textual abstraction. In contrast to event log abstractions, it is more difficult to enforce the
context window limit because of missing criteria to rank the importance of the elements of the process model. Textual abstractions of large Petri nets or declarative models with many rules can be provided only to LLMs with a large context window.

\begin{table*}[!t]
\centering
\resizebox{\textwidth}{!}{
\begin{tabular}{ll}
\toprule
\textbf{Command} & \textbf{Description} \\
\midrule
\texttt{pm4py.vis.save\_vis\_petri\_net} & Saves the visualization of a Petri net model. \\
\texttt{pm4py.vis.save\_vis\_dfg} & Saves the visualization of a directly-follows graph annotated with the frequency. \\
\texttt{pm4py.vis.save\_vis\_performance\_dfg} & Saves the visualization of a directly-follows graph annotated with the performance. \\
\texttt{pm4py.vis.save\_vis\_process\_tree} & Saves the visualization of a process tree. \\
\texttt{pm4py.vis.save\_vis\_bpmn} & Saves the visualization of a BPMN model. \\
\texttt{pm4py.vis.save\_vis\_heuristics\_net} & Saves the visualization of an heuristics net. \\
\texttt{pm4py.vis.save\_vis\_dotted\_chart} & Saves the visualization of a dotted chart. \\
\texttt{pm4py.vis.save\_vis\_sna} & Saves the visualization of the results of a social network analysis. \\
\texttt{pm4py.vis.save\_vis\_case\_duration\_graph} & Saves the visualization of the case duration graph. \\
\texttt{pm4py.vis.save\_vis\_events\_per\_time\_graph} & Saves the visualization of the events per time graph. \\
\texttt{pm4py.vis.save\_vis\_performance\_spectrum} & Saves the visualization of the performance spectrum. \\
\texttt{pm4py.vis.save\_vis\_events\_distribution\_graph} & Saves the visualization of the events distribution graph. \\
\texttt{pm4py.vis.save\_vis\_ocdfg} & Saves the visualization of an object-centric directly-follows graph. \\
\texttt{pm4py.vis.save\_vis\_ocpn} & Saves the visualization of an object-centric Petri net. \\
\texttt{pm4py.vis.save\_vis\_object\_graph} & Saves the visualization of an object-based graph. \\
\texttt{pm4py.vis.save\_vis\_network\_analysis} & Saves the visualization of the results of a network analysis. \\
\texttt{pm4py.vis.save\_vis\_transition\_system} & Saves the visualization of the results of a transition system. \\
\texttt{pm4py.vis.save\_vis\_prefix\_tree} & Saves the visualization of a prefix tree. \\
\texttt{pm4py.vis.save\_vis\_alignments} & Saves the visualization of the alignments table. \\
\texttt{pm4py.vis.save\_vis\_footprints} & Saves the visualization of the footprints table. \\
\texttt{pm4py.vis.save\_vis\_powl} & Saves the visualization of a POWL model. \\
\bottomrule
\end{tabular}
}
\caption{Visualization methods provided by \emph{pm4py}.}
\label{table:pm4py_vis_commands}
\vspace{-4mm}
\end{table*}

\vspace{-2mm}
\subsection{Visual Abstractions of Process Mining Artifacts}
\label{subsec:visualAbstractionProcMinArt}

Visual process representations can also be a source of knowledge for PM on LVLMs.
The method
$$\mathtt{pm4py.llm.explain\_visualization(vis\_saver, *args, connector, **kwargs)}$$
can be used to provide a visualization to an LVMs,
in which $\mathtt{vis\_saver}$ is one of the visualizations described in Table \ref{table:pm4py_vis_commands},
$\mathtt{*args}$ (positional arguments) and $\mathtt{**kwargs}$ (keyword arguments) are the arguments provided to the visualization method,
and $\mathtt{connector}$ is the LVM that should be used. For instance, $\mathtt{pm4py.llm.openai\_query}$ can be used to query the \emph{gpt-4-vision-preview} model.
Several visualizations offer insights that are more immediate in comparison to the textual abstractions, for instance:
\begin{itemize}
\item Dotted charts show events and detect patterns like batching and concept drifts.
\item Performance spectrum visualizes activity flows over time, offering performance insights.
\item Case duration and events per time graphs help understand throughput and identify concept drifts.
\item Petri nets, BPMN, directly-follows graphs, and social network analysis benefit from visual layout for pattern detection.
\end{itemize}
It is important to acknowledge that prompts to LVLMs require more computationally intensive operations than textual prompts to LLMs. Moreover, LVLMs are not specialized in patterns that might be important for process mining, as highlighted in
\cite{DBLP:conf/bpm/Berti0A24}. Moreover, LVLMs may miss domain knowledge needed to interpret such visualizations (for instance, additional explanation may be required in the prompt to detect some patterns on the dotted charts).

\vspace{-2mm}
\subsection{SQL Queries against \emph{pm4py} Event Logs}
\label{subsec:sqlQueries}

Statements in natural language can be translated by LLMs in the most important SQL dialects \cite{DBLP:conf/nips/LiHQYLLWQGHZ0LC23}.
The data structure used for traditional event logs in \emph{pm4py} is the Pandas dataframe, which is a columnar data structure easily imported/exported from/to CSV and Parquet files.
Pandas dataframes can be queried in-memory using, for instance, the DuckDB SQL dialect.
The LLM needs to be provided with the attributes of the Pandas dataframe and their specification (case ID, activity, timestamp).
Moreover, if the target LLM's knowledge of DuckDB is not proficient, additional knowledge related to the database operators and process mining
need to be provided.
The SQL query can be executed using \texttt{duckdb.sql(sqlquery).to\_df()}, which returns a Pandas dataframe containing the answer to the question of the user.
It is important to note that only quantitative statements can be translated to SQL queries using this paradigm.
Also, the object-centric data structure in \emph{pm4py} is a collection of Pandas dataframes
(there are dataframes for the events, objects, and event-to-object relationships).
Therefore, SQL queries can be orchestrated against any combination of the dataframes part of the object-centric data structure,
enabling LLM-based querying of object-centric event logs\footnote{A textual description of the object-centric data structure in \emph{pm4py} is available at \url{https://github.com/fit-alessandro-berti/pm-manuals-for-llms/blob/main/pm4py_manual.txt}.}

\vspace{-2mm}
\subsection{Generating \emph{pm4py} Code with LLMs}
\label{subsec:generatingPm4pyCodeLLMs}

\emph{pm4py} comes with a rich documentation covering its features\footnote{The documentation is available at the URL \url{https://pm4py.fit.fraunhofer.de/static/assets/api/2.7.11/index.html}}.
Using \emph{RAG} or \emph{fine-tuning}, an LLM can be trained over the API of the library\footnote{A textual file collecting the APIs is available at the address \url{https://github.com/fit-alessandro-berti/pm-manuals-for-llms/blob/main/pm4py_manual.txt}.},
becoming able to translate user statements in executable Python code. This paradigm allows not only to query the event data but also to apply the algorithms provided by \emph{pm4py} and from any scientific Python library in general.
However, it might pose a security risk as code generated by the LLM could contain malicious instructions \cite{DBLP:journals/corr/abs-2312-02003}.
An example of LLM trained to generate \emph{pm4py} code is provided in the OpenAI's GPT store\footnote{\url{https://chat.openai.com/g/g-EjtLrVyWH-pm4py-assistant}}.

\vspace{-2mm}
\subsection{Automatic Formulation of Hypotheses}
\label{subsec:automaticFormulationHypotheses}

Combining the paradigms described in Section \ref{subsec:textualAbstractionProcMinArt} and \ref{subsec:sqlQueries}, it is possible to generate some hypotheses over the process mining artifacts, in particular concerning event data.
Usually, some information about the activities (for example, the textual abstraction of the directly-follows graph) and the attributes of the event log are provided along with a question to generate hypotheses.
The request can generically target any hypothesis over the event data or hypotheses over a specific domain.

An example script is provided for hypothesis generation\footnote{\url{https://github.com/pm4py/pm4py-core/blob/release/examples/llm/04_hypothesis_generation.py}}.
Each hypothesis comes with a description/explanation and an executable (DuckDB) SQL statement. After executing the statement, the result could be validated from the LLM.
If the hypothesis proves valid, then the process stops. Otherwise, refined hypotheses may be provided by the LLM.
The automatic formulation of hypotheses also suffers from LLM hallucinations \cite{DBLP:conf/esws/MartinoIT23} and, in a minor way, privacy issues due to the provision of some textual abstractions in the prompt requesting the
generation of hypotheses.

\section{Conclusion}
\label{sec:conclusion}

The \emph{pm4py} process mining library offers a rich integration of PM with LLM, implementing all the paradigms described in Section \ref{sec:implementationParadigms}.
Potential development directions, dependent on the future capabilities of LLMs/LVLMs include the extraction of event logs from videos, the generation of visual explanations for a business process,
and autonomous process discovery/conformance checking from the event data.

\bibliographystyle{splncs04}
\bibliography{references}

\end{document}